\documentclass[prc,superscriptaddress,unsortedaddress,twocolumn,showpacs]{revtex4}

\usepackage{graphicx}
\usepackage{dcolumn}
\usepackage{bm}
\usepackage{enumerate}

\usepackage{amsmath}
\usepackage{amssymb}
\usepackage{times}

\usepackage{color}

\newcommand{\beq}{\begin{equation}}
\newcommand{\eeq}{\end{equation}}
\newcommand{\bea}{\begin{eqnarray}}
\newcommand{\eea}{\end{eqnarray}}

\def\fun#1#2{\lower3.6pt\vbox{\baselineskip0pt\lineskip.9pt
 \ialign{$\mathsurround=0pt#1\hfil##\hfil$\crcr#2\crcr\sim\crcr}}}

\pacs{24.10.-i, 25.60.Gc, 21.10.Gv, 27.20.+n}

\begin{document}

\title{
Analysis of a low-energy correction to the eikonal approximation
}

\author{Tokuro Fukui}
\email{tokuro@rcnp.osaka-u.ac.jp}
\affiliation{
Research Center for Nuclear Physics, Osaka University, Ibaraki, Osaka 567-0047, Japan
}

\author{Kazuyuki Ogata}
\email{kazuyuki@rcnp.osaka-u.ac.jp}
\affiliation{
Research Center for Nuclear Physics, Osaka University, Ibaraki, Osaka 567-0047, Japan
}

\author{Pierre Capel}
\email{pierre.capel@ulb.ac.be}
\affiliation{
Physique Nucl\'{e}aire et Physique Quantique C.P. 229, Universit\'{e} Libre de Bruxelles
(ULB), B-1050 Brussels, Belgium
}

\date{\today}

\begin{abstract}
Extensions of the eikonal approximation to low energy
(20~MeV/nucleon typically) are studied.
The relation between the dynamical eikonal approximation (DEA) and
the continuum-discretized coupled-channels method with the eikonal
approximation (E-CDCC) is discussed.
When Coulomb interaction is artificially turned off, DEA and E-CDCC
are shown to give the same breakup cross section, within 3\% error,
of $^{15}$C on $^{208}$Pb at 20~MeV/nucleon.
When the Coulomb interaction is included, the difference is appreciable
and none of these models agrees with full CDCC calculations.
An empirical correction significantly reduces this difference.
In addition, E-CDCC has a convergence problem. By including a quantum-mechanical
correction to E-CDCC for lower partial waves between $^{15}$C and
$^{208}$Pb, this problem is resolved and the result perfectly reproduces
full CDCC calculations at a lower computational cost.
\end{abstract}

\maketitle

\section{Introduction}
\label{s1}

The development of radioactive-ion beams in the mid-1980s has enabled
the exploration of the nuclear landscape far from stability.
This technical breakthrough has led to the discovery of exotic nuclear
structures, like nuclear halos and shell inversions.
Halo nuclei exhibit a very large matter radius compared to their isobars.
This unusual feature is explained by a strongly clusterized structure:
a compact core that contains most of the nucleons to which one or
two neutrons are loosely bound.
Due to quantum tunneling, these valence neutrons exhibit a large
probability of presence at a large distance from the core,
hence increasing significantly the radius of the nucleus.
Examples of one-neutron halo nuclei are $^{11}$Be and $^{15}$C,
while $^{6}$He and $^{11}$Li exhibit two neutrons in their halo.
Though less probable, proton halos are also possible.
The exotic halo structure has thus been the subject of many theoretical and
experimental studies for the last 30 years \cite{Tan96,Jon04}.

Due to their very short lifetime, halo nuclei must be studied
through indirect techniques, such as reactions.
The most widely used reaction to study halo nuclei is the breakup
reaction, in which the halo dissociates from the core through
the interaction with a target.
The extraction of reliable structure information from measurements
requires a good understanding of the reaction process.
Various models have been developed to describe the breakup of
two-body projectiles, i.e. one-nucleon halo nuclei
(see Ref.~\cite{BC12} for a review).

The continuum-discretized coupled channel method (CDCC) is
a fully quantum model in which the wave function describing
the three-body motion---two-body projectile plus target---is
expanded over the projectile eigenstates \cite{Kam86,Aus87,Yah12}.
For breakup modeling, the core-halo continuum must be included
and hence is discretized and truncated to form an approximate
complete set of states.
With such an expansion, the corresponding Schr\"odinger equation
translates into a set of coupled equations~\cite{Yah12,Aus89,Aus96}.
This reaction model is very general and has been successfully used
to describe several real and virtual breakup reactions at both low
and intermediate energies~\cite{Mor01,Mat03,Ega04,Man05}.
However it can be very computationally challenging,
especially at high beam energy.
This hinders the extension of CDCC models to reactions beyond
the simple usual two-body description of the projectile.
Simplifying approximations, less computationally demanding,
can provide an efficient way to avoid that limitation of CDCC.

At sufficiently high energy, the eikonal approximation can be performed.
In that approximation, the projectile-target relative motion is assumed
not to deviate significantly from the asymptotic plane wave~\cite{Gla59}.
By factorizing that plane wave out of the three-body wave function,
the Schr\"odinger equation can be significantly simplified.
Both the eikonal CDCC (E-CDCC)~\cite{Oga03,Oga06} and the dynamical
eikonal approximation (DEA)~\cite{Bay05,Gol06} are such eikonal models.
Note that these models differ from the usual eikonal approximation in that
they do not include the subsequent adiabatic approximation,
in which the internal dynamics of the projectile is neglected \cite{Suz03}.

The E-CDCC model solves the eikonal equation using the same discretization
technique as the full CDCC model.
Thanks to this, E-CDCC can be easily extended to a hybrid version,
in which a quantum-mechanical (QM) correction to the scattering amplitude
can be included for the low orbital angular momentum $L$
between the projectile and the target.
This helps in obtaining results as accurate as a full CDCC with a minimal task.
In addition, E-CDCC can take the dynamical relativistic effects into
account~\cite{OB09,OB10}, and it has recently been extended to
inclusive breakup processes~\cite{Yah11,Has11}.

Within the DEA, the eikonal equation is solved by expanding the projectile wave function
upon a three-dimensional mesh, i.e., without the CDCC partial-wave expansion \cite{Bay05}.
This prescription is expected to efficiently include components of the projectile
wave function up to high orbital angular momentum between its constituents.
Moreover it enables describing both bound and breakup states on the same footing,
without resorting to continuum discretization.
Since DEA treats the three-body dynamics explicitly,
all coupled-channels effects are automatically included.
Excellent agreement with the experiment has been obtained at the DEA for
the breakup and elastic scattering of one-nucleon halo nuclei on both heavy
and light targets at intermediate energies (e.g., 70~MeV/nucleon) \cite{Gol06}.

In a recent work~\cite{Cap12}, a comparison between
CDCC and DEA was performed.
The breakup of the one-neutron halo nucleus $^{15}$C
on $^{208}$Pb has been chosen as a test case.
At 68~MeV/nucleon, the results of the two models agree
very well with each other.
At 20~MeV/nucleon, DEA cannot reproduce the CDCC results
because the eikonal approximation is no longer valid at such low energy.
It appears that the problem is due to the Coulomb deflection,
which, at low energy, significantly distorts the projectile-target
relative motion from a pure plane wave.
Because of the computational advantage of the eikonal approximation
over the CDCC framework, it is important to pin down where the difference
comes from in more detail and try to find a way to correct it.

The goal of the present paper is to analyze in detail the QM correction
to E-CDCC in its hybrid version to see if it can be incorporated within
the DEA to correct the lack of Coulomb deflection observed in that reaction model.
To do so,  we compare E-CDCC and DEA with a special emphasis upon
their treatment of the Coulomb interaction.
We focus on the $^{15}$C breakup on $^{208}$Pb at 20~MeV/nucleon.
First, we compare the results of DEA and E-CDCC with the
Coulomb interaction turned off to show that both models solve essentially
the same Schr\"odinger equation.
Then we include the Coulomb interaction
to confirm the Coulomb deflection effect observed by the authors of Ref.~\cite{Cap12}.
We check that the hybrid version of E-CDCC reproduces correctly the
full CDCC calculations, and analyze this QM correction to suggest
an approximation that can be implemented within DEA to simulate the
Coulomb deflection.

In Sec.~\ref{s2} we briefly review DEA and E-CDCC, and clarify the
relation between them. We compare in Sec.~\ref{s3} the breakup
cross sections of $^{15}$C on $^{208}$Pb at 20~MeV/nucleon, with
and without the Coulomb interaction. A summary is given in Sec.~\ref{s4}.

\section{Formalism}
\label{s2}

\subsection{Three-body reaction system}
\label{s2-1}

\begin{figure}[htbp]
\begin{center}
\includegraphics[width=6cm]{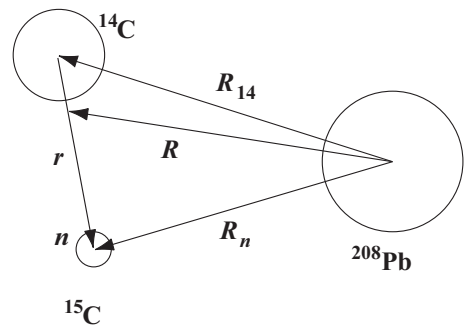}
\caption{Schematic illustration of the $({}^{14}{\rm C}+n)+{}^{208}{\rm Pb}$
three-body system.}
\label{fig1}
\end{center}
\end{figure}

We describe the $^{15}$C breakup on $^{208}$Pb using the coordinate
system shown in Fig.~\ref{fig1}. The coordinate of the center-of-mass (c.m.)
of $^{15}$C relative to $^{208}$Pb is denoted by ${\bm R}$,
and ${\bm r}$ is the neutron-$^{14}$C relative coordinate.
${\bm R}_{n}$ and ${\bm R}_{14}$ are, respectively,
the coordinates of neutron $n$ and the c.m. of $^{14}$C from $^{208}$Pb.
%
We assume both $^{14}$C and $^{208}$Pb to be inert nuclei.
In this study we neglect the spin of $n$.
The Hamiltonian describing the $^{15}$C structure therefore reads
\beq
h=-\frac{\hbar^2}{2\mu_{n14}}\Delta_{\bm r}+U_{nC}({\bm r}),
\label{eh}
\eeq
where $\mu_{n14}$ is the $^{14}$C-$n$ reduced mass and $U_{nC}$ is
a phenomenological potential describing the $^{14}$C-$n$ interaction
(see Sec.~\ref{s3-1}).
We denote by $\varphi_{\varepsilon \ell m}$ the eigenstates of Hamiltonian
of Eq.~(\ref{eh}) at energy $\varepsilon$ in partial wave $\ell m$, with
$\ell$ the $^{14}$C-$n$ orbital angular momentum and $m$ its projection.
For negative energies, these states are discrete and describe bound
states of the nucleus. For the present comparison, we consider the sole
ground state $\varphi_{0 \ell_0 m_0}$ at energy $\varepsilon_0=-1.218$~MeV.
The positive-energy eigenstates correspond to continuum states
that simulate the broken up projectile.

To simulate the interaction between $n$ ($ ^{14}$C) and $^{208}$Pb,
we adopt the optical potential $U_{n}$ ($U_{\rm 14}$) (see Sec.~\ref{s3-1}).
Within this framework, the study of $^{15}$C-$^{208}$Pb collision
reduces to solving the three-body Schr\"odinger equation
\beq
H\Psi({\bm R},{\bm r})=E_{\rm tot}\Psi({\bm R},{\bm r})
\label{e3b}
\eeq
with the Hamiltonian
\beq
H=-\frac{\hbar^2}{2\mu}\Delta_{\bm R}+h+U_{14}({\bm R}_{14})+U_n({\bm R}_{n}),
\label{eH3b}
\eeq
where $\mu$ is the $ ^{15}$C-$^{208}$Pb reduced mass.
Equation (\ref{e3b}) has to be solved with the incoming boundary condition
\beq
\lim_{z\rightarrow -\infty}\Psi({\bm R},{\bm r})=e^{iK_0z+\cdots}\varphi_{0 \ell_0 m_0}({\bm r}),
\label{einit}
\eeq
where $K_0$ is the wave number for the initial projectile-target motion,
whose direction defines the $z$ axis.
That wave number is related to the total energy
$E_{\rm tot}=\hbar^2K_0^2/(2\mu)+\varepsilon_0$.
The ``$\cdots$'' in Eq.~(\ref{einit}) indicates that the projectile-target
relative motion is distorted by the Coulomb interaction, even at large distances.

In the eikonal approximation, the three-body wave function $\Psi$ is
assumed not to vary significantly from the incoming plane wave of Eq.~(\ref{einit}).
Hence the usual eikonal factorization
\beq
\Psi({\bm R},{\bm r})=e^{iK_0z}\psi({\bm b},z,{\bm r}),
\label{efact}
\eeq
where we have explicitly decomposed ${\bm R}$ in its longitudinal $z$
and transverse ${\bm b}$ components.
In the following ${\bm b}$ is expressed as ${\bm b}=(b,\phi_R)$ with
$b$ the impact parameter and $\phi_R$ the azimuthal angle of ${\bm R}$.

Using factorization Eq.~(\ref{efact}) in Eq.~(\ref{e3b}) and
taking into account that $\psi$ varies smoothly with ${\bm R}$,
we obtain equations simpler to solve than the full three-body
Schr\"odinger equation (\ref{e3b}).
In the following subsections, we specify the equations solved within
the E-CDCC (Sec.~\ref{s2-2}) and the DEA (Sec.~\ref{s2-3}).

\subsection{Continuum-discretized coupled-channels method with the eikonal
approximation (E-CDCC)}
\label{s2-2}

E-CDCC expresses the three-body wave function $\Psi$ as \cite{Oga03,Oga06}
\beq
\Psi({\bm R},{\bm r})
=
\sum_{i \ell m}
\bar{\xi}_{i \ell m}(b,z)  \varphi_{i \ell m} ({\bm r})
e^{iK_i z}
e^{i(m_0-m)\phi_R}
\phi^{\rm C}_i (R),
\label{ecdcc}
\eeq
where $\{ \varphi_{i \ell m} \}$ are square-integrable states that describe
the eigenstates of Hamiltonian $h$ Eq.~(\ref{eh}).
The subscript $i$ labels the eigenenergy of $^{15}$C.
The ground state corresponds to $i=0$ and the corresponding wave function
$\varphi_{0 \ell_0 m_0}$ is the exact eigentstate of $h$.
The values $i>0$ correspond to discrete states simulating $^{15}$C continuum.
They are obtained by the binning technique \cite{Yah86}, viz by averaging exact
continuum states over small positive-energy intervals, or bins.
The set $\{ \varphi_{i \ell m} \}$ satisfying
\beq
\left\langle \varphi_{i' \ell' m'}
\left\vert
h
\right\vert
\varphi_{i \ell m} \right\rangle_{\bm r}
=
\varepsilon_i \delta_{i'i}\delta_{\ell'\ell}\delta_{m'm}
\eeq
is assumed to form an approximate complete set
for the projectile internal coordinate $\bm r$.
The plane wave $e^{iK_i z}$ contains the dominant part of
the projectile-target motion as explained above.
The corresponding wave number varies with the energy of the
eigenstate of $^{15}$C respecting the conservation of energy
$E_{\rm tot}=\hbar^2K_i^2/2\mu+\varepsilon_i$.
In Eq.~(\ref{ecdcc}), $\phi^{\rm C}_i$ is the approximate
Coulomb incident wave function given by
\beq
\phi^{\rm C}_i (R)=e^{i\eta_{i}\ln\left[K_{i}R-K_{i}z\right]}
\label{Coulwf}
\eeq
where $\eta_i$ is the Sommerfeld parameter corresponding to
the $i$th state of $^{15}$C
\beq
\eta_i = \frac{Z_{\rm C} Z_{\rm T}e^2\mu}{\hbar^2 K_i},
\label{eeta}
\eeq
with $Z_{\rm C}e$ ($Z_{\rm T}e$) the charge of the projectile (target).
The functions $\bar{\xi}$ are the coefficients of the expansion Eq.~(\ref{ecdcc})
that have to be evaluated numerically.

Inserting Eq.~(\ref{ecdcc}) into Eq.~(\ref{e3b}),
multiplied by $\varphi_{i' \ell' m'}$ on the left,
and integrating over ${\bm r}$, one gets \cite{Oga03,Oga06}
\bea
\lefteqn{\frac{\partial}{\partial z}\bar{\xi}_{c}(b,z)=}\nonumber\\
& &
\frac{1}{i\hbar v_i(R)}\sum_{c'}
\mathcal{F}_{cc'}(b,z)
\bar{\xi}_{c'}(b,z)
e^{i(K_{i'}-K_i)z}
\mathcal{R}_{ii^{\prime}}(b,z),\nonumber\\
\label{ECDCCeq}
\eea
where the index $c$ denotes $i$, $\ell$, and $m$ together.
In E-CDCC, the projectile-target velocity $v_i$ depends
on both the projectile excitation energy and its position following
\begin{equation}
v_{i} (R)  =\frac{1}{\mu}
\sqrt{  \hbar^2  K_{i}^2 - 2\mu V_{\rm C}(R)},
\label{vr}
\end{equation}
where
\beq
V_{\rm C}(R)=\frac{Z_{\rm C}Z_{\rm T}e^{2}}{R}
\label{eVC}
\eeq
is the projectile-target potential that slows down the projectile
as it approaches the target.
The coupling potential $\mathcal{F}_{cc'}$ is defined by
\beq
\mathcal{F}_{cc'}(b,z)
=
\left\langle \varphi_{c}
\left\vert
U_{14}+U_n-V_{\rm C}
\right\vert
\varphi_{c'} \right\rangle_{\bm r}
e^{i(m-m')\phi_R},
\label{ff}
\eeq
and
\begin{equation}
\mathcal{R}_{ii^{\prime}}(b,z)
=
\frac{\left(K_{i'}R-K_{i'}z\right)  ^{i\eta_{i'}}}
{\left(K_{i}R-K_{i}z\right)  ^{i\eta_{i}}}.
\label{Rfac}
\end{equation}
Within the E-CDCC framework, the boundary condition Eq.~(\ref{einit})
translates into
\beq
\lim_{z\to -\infty}\bar{\xi}_{i \ell m}(b,z)
=
\delta_{i 0} \delta_{\ell \ell_0} \delta_{m m_0}.
\eeq

\subsection{The dynamical eikonal approximation (DEA)}
\label{s2-3}

In the DEA, the three-body wave function is factorized following \cite{Bay05,Gol06}
\beq
\Psi({\bm R},{\bm r})
=
\psi\left(  \boldsymbol{b},z,\boldsymbol{r}\right)
e^{iK_0 z}
e^{i\chi_{\rm C}(b,z)}
e^{i \varepsilon_0 z / ( \hbar v_0)},
\label{DEAwf}
\eeq
where $\chi_{\rm C}$ is the Coulomb phase that accounts for the
Coulomb projectile-target scattering
\beq
\chi_{\rm C}(b,z)=
-\frac{1}{\hbar v_0}\int_{-\infty}^{z} V_{\rm C}(R) \; dz',
\label{echiC}
\eeq
where $v_0=\hbar K_0/\mu$ is the initial velocity of the projectile.
Note that the phase
$\exp{[\varepsilon_0 z / (i \hbar v_0)]}$
can be ignored as it has no effect on physical observables \cite{Gol06}.

From the factorization in Eq.~(\ref{DEAwf}), we obtain the DEA equation~\cite{Bay05,Gol06}
\beq
i\hbar v_0\frac{\partial}{\partial z}
\psi\left(  \boldsymbol{b},z,\boldsymbol{r}\right)=
\left[
h+ U_{\rm 14}+U_{n}-\varepsilon_{0}-V_{\rm C}
\right]
\psi\left(\boldsymbol{b},z,\boldsymbol{r}\right).
\label{DEAeq}
\eeq
The initial condition of Eq.~(\ref{einit}) translates into
\beq
\lim_{z \to -\infty} \psi\left(  \boldsymbol{b},z,\boldsymbol{r}\right)
=
\varphi_{0 \ell_0 m_0}({\bm r}).
\label{bc}
\eeq

The DEA equation~(\ref{DEAeq}) is solved for all ${\bm b}$
with respect to $z$ and ${\bm r}$ expanding the wave function
$\psi$ on a three-dimensional mesh.
This allows to include naturally all relevant states of $^{15}$C, i.e.,
eigenenergies $\varepsilon$ up to high values in the $n$-$^{14}$C continuum,
and large angular momentum $\ell$, and its $z$-component $m$.
This resolution is performed assuming a constant projectile-target
relative velocity $v=v_0$.
It should be noted that this does not mean
the adiabatic approximation, because in Eq.~(\ref{DEAeq}) the internal
Hamiltonian $h$ is explicitly included. The DEA thus treats properly the
change in the eigenenergy of $^{15}$C during the scattering process.
However, it does not change the $^{15}$C-$^{208}$Pb velocity accordingly.
This gives a violation of the conservation of the total energy of
the three-body system. However, even at 20~MeV/nucleon, its effect
is expected to be only a few percents as discussed below.

The calculation of physical observables requires the wave function
$\Psi$ of Eq.~(\ref{DEAwf}) at $z\to \infty$ \cite{Bay05,Gol06}.
The corresponding Coulomb phase $\chi_{\rm C}$ reads \cite{BD04}
\beq
\lim_{z\to \infty}\chi_{\rm C}=2 \eta_0 \ln (K_0 b),
\label{echiCinf}
\eeq
where $\eta_0$ is the Sommerfeld parameter for the entrance channel [see Eq.~(\ref{eeta})].

\subsection{Comparison between E-CDCC and DEA}
\label{s2-4}

To ease the comparison between the DEA and the E-CDCC,
we rewrite the formulas given in Sec.~\ref{s2-3}
in a coupled-channel representation.
We expand $\psi$ as
\beq
\psi\left(  \boldsymbol{b},z,\boldsymbol{r}\right)=
\sum_{i \ell m}
\xi_{i \ell m}(b,z)  \varphi_{i \ell m} ({\bm r})
e^{\varepsilon_{i}z /(i\hbar v_0)}
e^{i(m_0-m)\phi_R}.
\label{expansion}
\eeq

Inserting Eq.~(\ref{expansion}) into Eq.~(\ref{DEAeq}),
multiplied by $\varphi_{i' \ell' m'}$ from the left,
and integrating over ${\bm r}$, one gets
\beq
\frac{\partial}{\partial z}\xi_{c}(b,z)
=
\frac{1}{i\hbar v_0}\sum_{c'}
\mathcal{F}_{cc'}(b,z)
\xi_{c'}(b,z)
e^{(\varepsilon_{i'}-\varepsilon_{i})z/(i\hbar v_0)},
\label{DEAcc}
\eeq
which is nothing but the DEA equation (\ref{DEAeq}) in its coupled-channel
representation.

The boundary condition Eq.~(\ref{bc}) thus reads
\beq
\lim_{z\to -\infty}\xi_{i \ell m}(b,z)
=
\delta_{i 0} \delta_{\ell \ell_0} \delta_{m m_0}.
\eeq
Considering the expansion Eq.~(\ref{expansion}) in the
DEA factorization Eq.~(\ref{DEAwf}), the total wave function reads
\bea
\Psi({\bm R},{\bm r})
&=&
\sum_{c}
\xi_{c}(b,z)  \varphi_{c} ({\bm r})
e^{(\varepsilon_{i}-\varepsilon_{0})z/(i\hbar v_0)}
\nonumber \\
&&
\times e^{i(m_0-m)\phi_R}
e^{iK_0 z}
e^{i\chi_{\rm C}(b,z)}.
\label{DEAwfcc}
\eea

One may summarize the difference between Eqs.~(\ref{DEAcc})
and~(\ref{ECDCCeq}) as follows.
First, the DEA uses the constant and channel-independent
$^{15}$C-$^{208}$Pb relative velocity $v_0$,
whereas E-CDCC uses the velocity depending on both $R$
and the channel $i$ that ensures the total-energy conservation.

Second, whereas the right-hand side of Eq.~(\ref {DEAcc}) involves the phase
$\exp{[(\varepsilon_{i'}-\varepsilon_i) z / (i\hbar v_0)]}$, the E-CDCC
Eq.~(\ref{ECDCCeq}) includes the phase
$\exp{[i(K_{i'}-K_i)z]}$.
The former can be rewritten as
\beq
\frac{ \varepsilon_{i'}-\varepsilon_{i} }{i\hbar v_0}z
=
\frac{\hbar^{2}(K_{i}^{2}-K_{i'}^{2}) \; \mu z}
{2\mu \; i\hbar^{2}K_{0}}
=
\frac{K_{i'}+K_{i}}{2K_{0}}i\left(  K_{i'}-K_{i}\right)  z.
\label{exponent}
\eeq
If we can assume the semi-adiabatic approximation
\begin{equation}
\frac{K_{i'}+K_{i}}{2K_{0}}\approx 1,
\label{semiad}
\end{equation}
the exponent Eq.~(\ref{exponent}) becomes the same as in E-CDCC.
In the model space taken in the present study,
Eq.~(\ref{semiad}) holds within 1.5\% error
at 20~MeV/nucleon of incident energy.

Third, E-CDCC equation
contains $\mathcal{R}_{ii^{\prime}}$ taking account of
the channel dependence of the $^{15}$C-$^{208}$Pb Coulomb
wave function, which DEA neglects.
Nevertheless, it should be noted that, as shown in Refs.~\cite{Oga03,Oga06},
the Coulomb wave functions in the initial and final channels
involved in the transition matrix ($T$ matrix) of E-CDCC
eventually give a phase $2 \eta_j \ln (K_j b)$,
with $j$ the energy index in the final channel.
Thus, if Eq.~(\ref{semiad}) holds, the role of the Coulomb
wave function in the evaluation of the $T$ matrix in E-CDCC is expected
to be the same as in DEA, since DEA explicitly includes the Coulomb eikonal
phase, Eq.~(\ref{echiCinf}).

When the Coulomb interaction is absent, we have
$\mathcal{R}_{ii^{\prime}}(b,z)=1$ and no $R$ dependence of the velocity.
Therefore, it will be interesting to compare
the results of DEA and E-CDCC with and without the Coulomb interaction
separately.

\section{Results and discussion}
\label{s3}

\subsection{Model setting}
\label{s3-1}

We calculate the energy spectrum $d\sigma/d\varepsilon$
and the angular distribution $d\sigma/d\Omega$ of the
breakup cross section of $^{15}$C on $^{208}$Pb at 20~MeV/nucleon, where
$\varepsilon$ is the relative energy between $n$ and $^{14}$C after
breakup, and $\Omega$ is the scattering angle of the c.m. of the
$n$-$^{14}$C system.
We use the potential parameters shown in Table~\ref{tab1} for
$U_{n{\rm C}}$ (the $n$-$^{14}$C interaction), $U_{14}$, and $U_{n}$ \cite{Cap12};
the depth of $U_{n{\rm C}}$ for the $d$-wave is changed to
69.43~MeV to avoid a non-physical $d$ resonance.
The spin of the neutron is disregarded as mentioned earlier.
We adopt Woods-Saxon potentials for the interactions:
\bea
U_x (R_x)
&=&
-V_0 f(R_x,R_0,a_0)
-iW_v f(R_x,R_w,a_w)
\nonumber \\
&&+iW_s \dfrac{d}{dR_x}f(R_x,R_w,a_w)
\eea
with $f(R_x,\alpha,\beta)=(1+\exp[(R_x-\alpha)/\beta])^{-1}$;
$R_x=r$, $R_{14}$, and $R_n$ for $x=n{\rm C}$, $14$, and $n$ ,
respectively.
The Coulomb interaction between $^{14}$C and $^{208}$Pb is described
by assuming a uniformly charged sphere of radius $R_{\rm C}$.
%
\begin{table}[htb]
\caption{Potential parameters for the pair interactions
$U_{n{\rm C}}$, $U_{14}$, and $U_{n}$ \cite{Cap12}.
}
\begin{tabular}{cccccccccc}
\hline \hline
  & $V_0$ & $R_0$ & $a_0$ & $W_v$ & $W_s$ & $R_w$ & $a_w$ & $R_{\rm C}$ \\
  & (MeV) & (fm)  & (fm)  & (MeV) & (MeV) & (fm)  & (fm)  & (fm)        \\
\hline
$U_{n{\rm C}}$
  & 63.02 & 2.651 & 0.600 & ---   & ---   & ---   & ---   & ---         \\
$U_{14}$
  & 50.00 & 9.844 & 0.682 & 50.00 & ---   & 9.544 & 0.682 & 10.84       \\
$U_{n}$
  & 44.82 & 6.932 & 0.750 & 2.840 & 21.85 & 7.466 & 0.580 & ---         \\
\hline \hline
\end{tabular}
\label{tab1}
\end{table}

Unless stated otherwise, the model spaces chosen for our calculations
give a confidence level better than 3\% on the cross sections presented
in Secs.~\ref{s3-2} and \ref{s3-3}.
In E-CDCC, we take the maximum value of $r$ to be 800~fm with the
increment of 0.2~fm. When the Coulomb interaction is turned off,
we take the $n$-$^{14}$C partial waves up to $\ell_{\rm max}=10$.
For each $\ell$ the continuum state is truncated at
$k_{\max}=1.4$~fm$^{-1}$ and discretized into 35 states with the equal
spacing of $\Delta k=0.04$~fm$^{-1}$; $k$ is the relative wave number
between $n$ and $^{14}$C. The resulting number of coupled channels,
$N_{\rm ch}$, is 2311.
The maximum values of $z$ and $b$, $z_{\rm max}$ and $b_{\rm max}$,
respectively, are both set to 50~fm.
When the Coulomb interaction is included, we use $\ell_{\rm max}=6$,
$k_{\max}=0.84$~fm$^{-1}$, $\Delta k=0.04$~fm$^{-1}$,
$z_{\rm max}=1000$~fm, and $b_{\rm max}=150$~fm.
We have $N_{\rm ch}=589$ in this case.

In the DEA calculations, we use the same numerical parameters as in Ref.~\cite{Cap12}.
In the purely nuclear case,
the wave function $\psi$ is expanded over an angular mesh
containing up to $N_\theta\times N_\varphi=14\times27$ points,
a quasi-uniform radial mesh that extends up to 200~fm with 200 points,
$b_{\max}=50$~fm, and $z_{\max}=200$~fm (see Ref.~\cite{Cap03} for details).
In the charged case,
the angular mesh contains up to $N_\theta\times N_\varphi=12\times23$ points,
the radial mesh extends up to 800~fm with 800 points,
$b_{\max}=300$~fm, and $z_{\max}=800$~fm.

\subsection{Comparison without Coulomb interaction}
\label{s3-2}

%
\begin{figure}[b]
\begin{center}
\includegraphics[width=7.5cm]{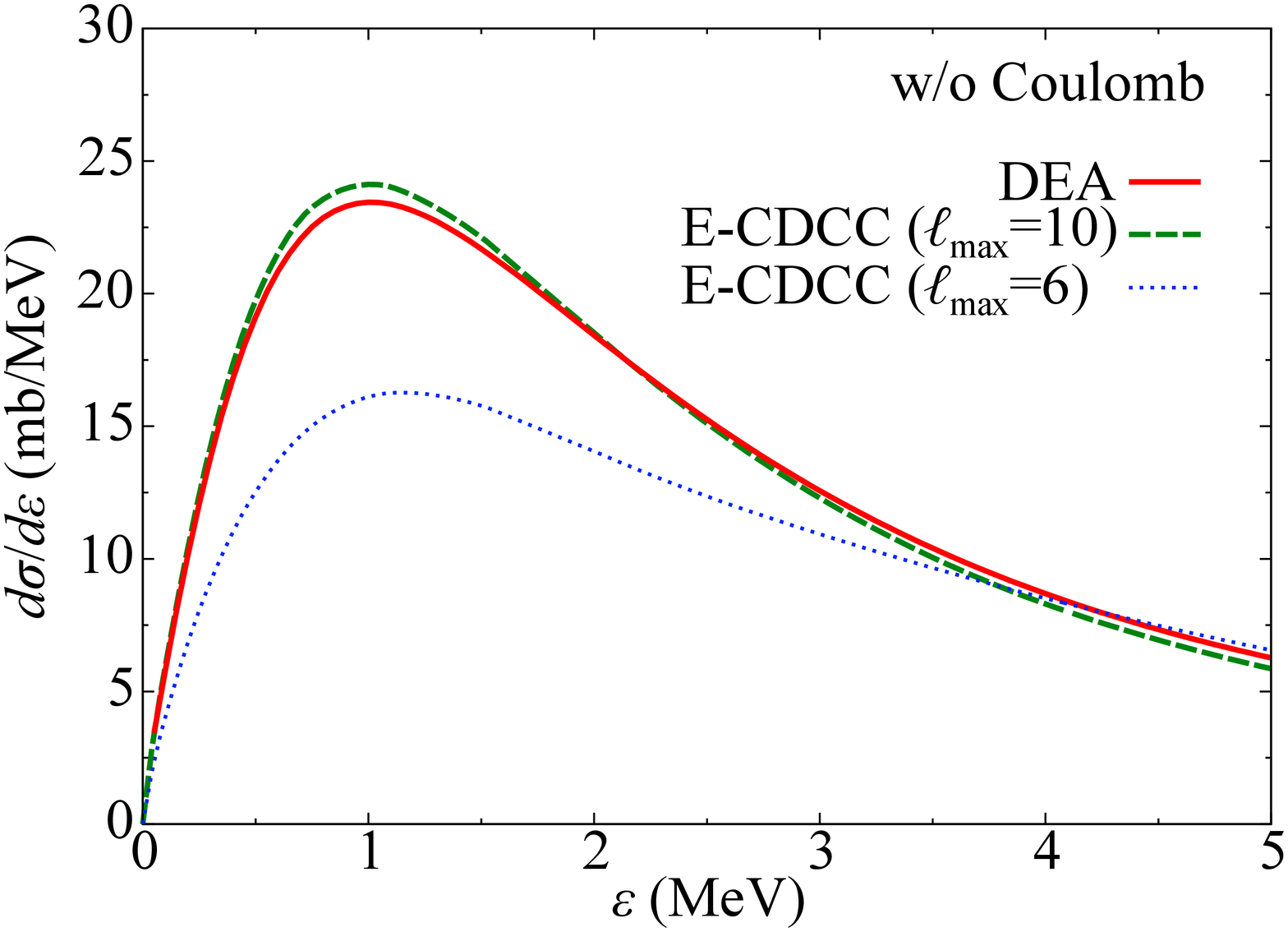}
\caption{(Color online) Energy spectrum of the $^{15}$C breakup cross section
on $^{208}$Pb at 20~MeV/nucleon with the Coulomb interaction turned off.
The solid and dashed lines show the results obtained by DEA and E-CDCC,
respectively. The result of E-CDCC with $\ell_{\rm max}=6$ is denoted by
the dotted line.
}
\label{fig2}
\end{center}
\end{figure}
The goal of the present work being to study the difference in
the treatment of the Coulomb breakup between the E-CDCC and DEA,
we first check that both models agree when the Coulomb interaction is switched off.
We show in Fig.~\ref{fig2} the results of $d\sigma/d\varepsilon$
calculated by DEA (solid line) and E-CDCC (dashed line);
$d\sigma/d\varepsilon$
is obtained by integrating the double-differential breakup
cross section $d^2\sigma/(d\varepsilon d\Omega)$
over $\Omega$ in the whole variable region.
The two results agree very well with each other; the difference
around the peak is below 3\%.

\begin{figure}[htbp]
\begin{center}
\includegraphics[width=7.5cm]{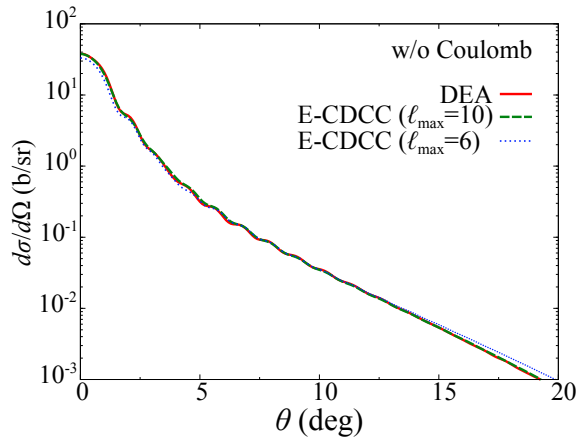}
\caption{(Color online) Same as Fig.~\ref{fig2} but for the angular distribution.
}
\label{fig3}
\end{center}
\end{figure}
In Fig.~\ref{fig3} the comparison in $d\sigma/d\Omega$,
i.e., $d^2\sigma/(d\varepsilon d\Omega)$ integrated over
$\varepsilon$ up to 10~MeV, is shown. The agreement between
the two models is excellent confirming that,
when the Coulomb interaction is turned off,
the DEA and E-CDCC solve the same equation and give the same
result, as expected from the discussion at the end of Sec.~\ref {s2-4}.
In particular this comparison shows that Eq.~(\ref{semiad})
turns out to be satisfied with very high accuracy.
It should be noted that the good agreement between the DEA and E-CDCC is
obtained only when a very large model space is taken. In fact,
if we put $\ell_{\rm max}=6$ in E-CDCC, we have 30\% smaller
$d\sigma/d\varepsilon$ (dotted line) than the converged value and, more
seriously, even the shape cannot be reproduced. This result shows
the importance of the higher partial waves of $n$-$^{14}$C for
the nuclear breakup at 20~MeV/nucleon.

\subsection{Comparison with Coulomb interaction}
\label{s3-3}

\begin{figure}[b]
\begin{center}
\includegraphics[width=7.5cm]{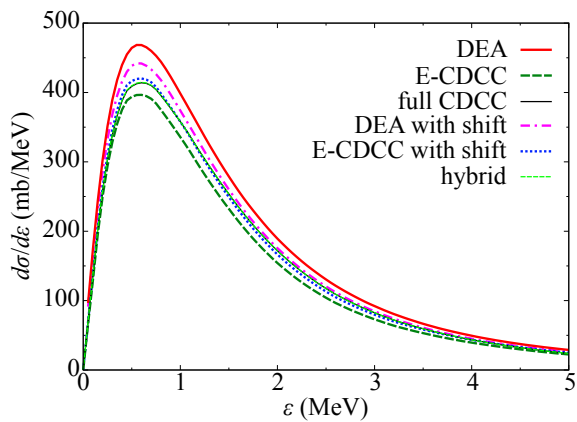}
\caption{(Color online) Energy spectrum of the $^{15}$C breakup cross section
on $^{208}$Pb at 20~MeV/nucleon including the Coulomb interaction.
The solid, dashed, and thin solid lines show the results obtained by
DEA, E-CDCC, and full (QM) CDCC, respectively.
The results obtained with the correction (\ref{ebp}) are displayed with a dash-dotted line
for DEA and a dotted line for E-CDCC.
The calculation using the QM correction of E-CDCC, i.e., the hybrid
calculation, is shown by the light-green thin dashed line (superimposed onto the thin solid line).}
\label{fig4}
\end{center}
\end{figure}

\begin{figure}[t]
\begin{center}
\includegraphics[width=7.5cm]{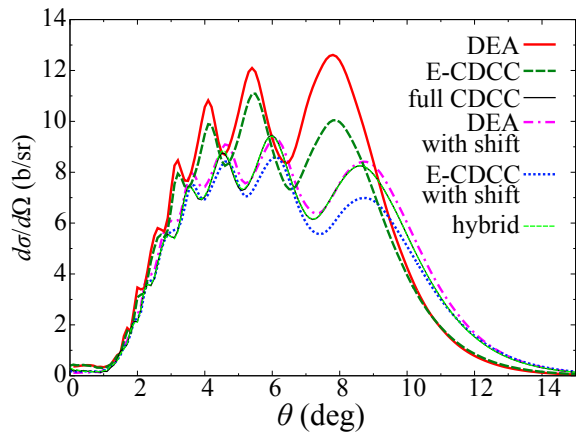}
\caption{(Color online) Same as Fig.~\ref{fig4} but for the angular distribution.
}
\label{fig5}
\end{center}
\end{figure}

\begin{figure}[b]
\begin{center}
\includegraphics[width=7.5cm]{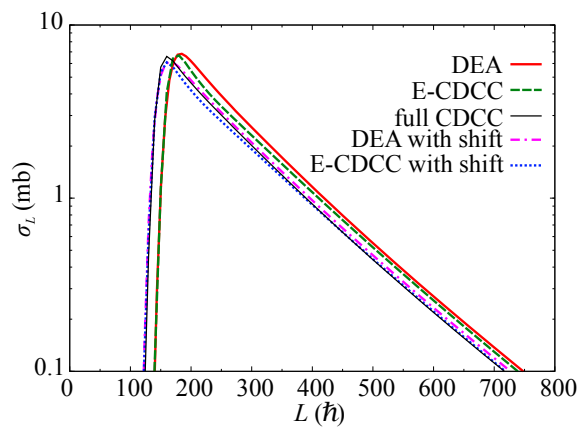}
\caption{(Color online) Contribution to the total breakup cross section
per projectile-target angular momentum $L$.
Neglecting the Coulomb deflection, DEA and E-CDCC are shifted to large $L$
compared to the full CDCC.
The correction Eq.~(\ref{ebp}) significantly reduces this shift for both models.}
\label{fig6}
\end{center}
\end{figure}

When the Coulomb interaction is switched on, DEA and E-CDCC
no longer agree with each other.
As seen in Fig.~\ref{fig4}, the DEA energy spectrum (solid line)
is much larger than the E-CDCC one (dashed line).
Moreover none of them agrees with the full CDCC calculation (thin solid line):
DEA is too high while E-CDCC is too low.
%
The discrepancy of both models with the fully quantal calculation
manifests itself even more clearly in the angular distribution.
In Fig.~\ref{fig5} we see that not only do the DEA and E-CDCC
cross sections differ in magnitude, but---as already seen in Ref.~\cite{Cap12}---their
oscillatory pattern is shifted to forward angle compared to the CDCC calculation.
%
To understand where the problem comes from we analyze in Fig.~\ref{fig6}
the contribution to the total breakup cross section of each
projectile-target relative angular momentum $L$ .
As expected from Figs.~\ref{fig4} and \ref{fig5}, the DEA calculation is larger than
the E-CDCC one, and this is observed over the whole $L$ range.
However, the most striking feature is to see that both models seem to be shifted
to larger $L$ compared to the full CDCC calculation.
To correct this, we replace in our calculations the transverse component of the
projectile-target relative coordinate $b$ by the empirical value \cite{BW81,Ber03,BD04}
\beq
b'=\frac{\eta_0}{K_0}+\sqrt{\frac{\eta_0^2}{K_0^2}+b^2}.
\label{ebp}
\eeq
The corresponding results are displayed in Figs.~\ref{fig4}, \ref{fig5} and \ref{fig6}
as dash-dotted lines for DEA and dotted lines for E-CDCC.

The correction Eq.~(\ref{ebp}) is very effective.
It significantly reduces the shift observed in the $L$ contributions to the breakup
cross section (see Fig.~\ref{fig6}).
Accordingly, it brings both DEA and E-CDCC  energy spectra closer to the full CDCC one
(see Fig.~\ref{fig4}).
Note that for this observable the correction seems better for E-CDCC than for DEA:
even with the shift, the latter still exhibits a non-negligible enhancement with respect to
CDCC at low energy $E$.
More spectacular is the correction of the shift in the angular distribution
observed in Ref.~\cite{Cap12} and in Fig.~\ref{fig5}.
In particular, the shifted DEA cross section is now very close to the CDCC one, but at forward angles, where DEA overestimates CDCC. 
Once shifted, E-CDCC still underestimates slightly the full CDCC calculation.
However, its oscillatory pattern is now in phase with that of the CDCC cross section,
which is a big achievement in itself.
This shows that the lack of Coulomb deflection observed in Ref.~\cite{Cap12} for
eikonal-based calculations can be efficiently corrected by the simple shift Eq.~(\ref{ebp})
suggested long ago \cite{BW81,BD04}.

Albeit efficient, the correction Eq.~(\ref{ebp}) is not perfect.
This is illustrated by the enhanced (shifted) DEA cross section observed in
the low-energy peak in Fig.~\ref{fig4}
and at forward angles in Fig.~\ref{fig5}.
Both problems can be related to the same root because the forward-angle part
of the angular distribution is dominated by low-energy contributions.
As shown in Ref.~\cite{Gol06}, that part of the cross section is itself
dominated by large $b$'s, at which the correction Eq.~(\ref{ebp}) is not fully sufficient.
As shown in Fig.~\ref{fig6}, the shifted DEA remains slightly larger than the full CDCC.
Future works may suggest a better way to handle this shift than the empirical
correction Eq.~(\ref{ebp}).
Nevertheless, these results show that this correction provides a simple, elegant, and
cost-effective way to account for Coulomb deflection in eikonal-based models.

\begin{figure}[tb]
\begin{center}
\includegraphics[width=7.5cm]{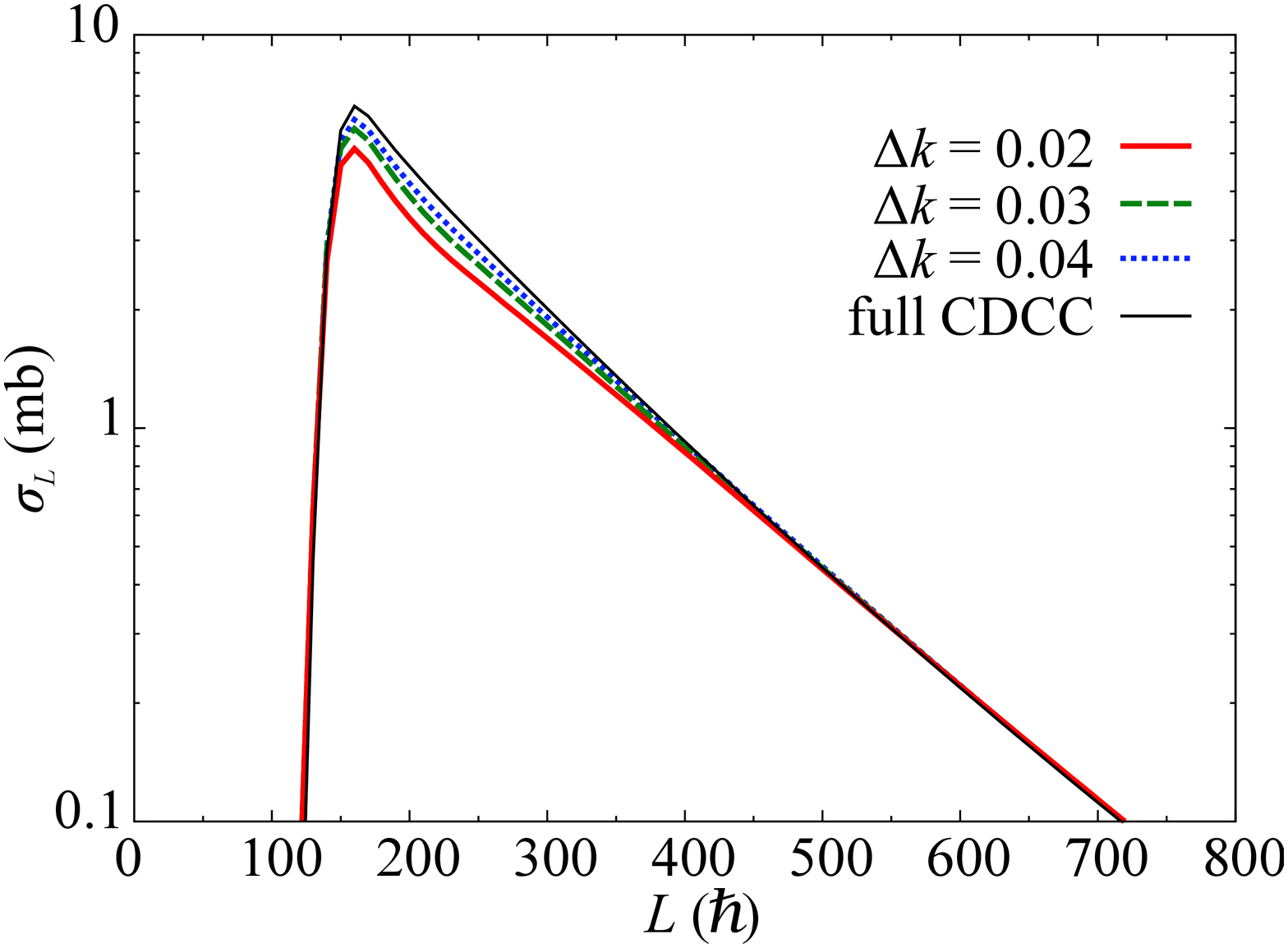}
\caption{(Color online) Convergence problem observed in (shifted) E-CDCC calculations:
cross sections computed with different bin widths do not converge towards the CDCC
calculation.}
\label{fig7}
\end{center}
\end{figure}

The underestimation of the full CDCC angular distribution by E-CDCC comes
most likely from a convergence problem within that reaction model.
This is illustrated in Fig.~\ref{fig7}, showing the $L$-contribution
to the total breakup cross section.
The thin solid line corresponds to the (converged) CDCC calculation,
whereas the other lines correspond to (shifted) E-CDCC calculations with bin widths
of $\Delta k=0.02$ (solid line), 0.03 (dashed line), and 0.04~fm$^{-1}$
(dotted line).
As can be seen, below $L\approx 500~\hbar$, no convergence can be obtained,
although CDCC has fully converged.
We cannot expect this model to provide accurate breakup cross sections.
The results displayed in Figs.~\ref{fig4} and \ref{fig5} are therefore unexpectedly good.
Note that the present ill-behavior of E-CDCC occurs only when
the Coulomb interaction involved is strong and the incident energy is
low; no such behavior was observed in previous
studies~\cite{Yah12,Oga03,Oga06,OB09,OB10}.
Interestingly, DEA does not exhibit such a convergence issue.
This is reminiscent of the work of Dasso \textit{et al.} \cite{Vit09},
where it was observed that reaction calculations converge faster
by expanding the wave function upon a mesh rather than
by discretization of the continuum.

The aforementioned results indicate that the shift Eq.~(\ref{ebp}) corrects
efficiently for the Coulomb deflection, which is expected to play a significant
role at large $L$.
At small $L$, we believe the nuclear projectile-target interaction will induce
significant couplings between various partial waves, which cannot be
accounted for by that simple correction.
To include these couplings, a hybrid solution between E-CDCC and
the full CDCC has been suggested \cite{Oga03,Oga06}.
At low $L$ a usual CDCC calculation is performed, which fully
accounts for the strong coupling expected from the nuclear
interaction between the projectile and the target.
At larger $L$, these couplings are expected to become
negligible, which implies that a (shifted) E-CDCC calculation should be reliable.
As explained in Refs.~\cite{Oga03,Oga06}, the transition
angular momentum $L_{\rm C}$ above which E-CDCC is used
is an additional parameter of the model space that has to be
determined in the convergence analysis.
Depending on the beam energy and the system studied,
usual values of $L_{\rm C}$ are in the range 400--$1000~\hbar$.
In the present case, due to the convergence issue observed in
E-CDCC, the value $L_{\rm C}=500~\hbar$ is chosen.

The quality of this hybrid solution is illustrated by the light-green thin dashed lines
in Figs.~\ref{fig4} and \ref{fig5}, which are barely visible as they are
superimposed to the full CDCC results.
The coupling of the hybrid solution to the Coulomb shift Eq.~(\ref{ebp})
enables us to reproduce exactly the CDCC calculations at a much
lower computational cost since
the computational time for each $b$ with E-CDCC is
about 1/60 of that for each $L$ with full CDCC.
In addition, it solves the convergence problem of E-CDCC.

\section{Summary}
\label{s4}

With the ultimate goal of understanding the ability of the hybrid
version of the continuum-discretized coupled-channels method with
the eikonal approximation (E-CDCC) to account for the discrepancy
between the dynamical eikonal approximation (DEA) and
the continuum discretized coupled channel model (CDCC)
observed in Ref.~\cite{Cap12}, we have
clarified the relation between the DEA and E-CDCC.
By using a coupled-channel representation of DEA equations,
DEA is shown to be formally equivalent to E-CDCC, if
the semi-adiabatic approximation of Eq.~(\ref{semiad}) is satisfied and
the Coulomb interaction is absent.

We consider the same test case as in Ref.~\cite{Cap12}, i.e.,
the breakup of $^{15}$C on $^{208}$Pb at 20~MeV/nucleon.
For this reaction Eq.~(\ref{semiad}) holds within 1.5\% error.
When the Coulomb interaction is artificially turned off,
DEA and E-CDCC
are found to give the same result within 3\% difference
for both the energy spectrum and the angular distribution.
This supports the equivalence of the two models for describing the
breakup process due purely to nuclear interactions.

Next we make a comparison including the Coulomb interaction.
In this case, DEA and E-CDCC no longer agree with each other and
they both disagree with the full CDCC calculation.
In particular both angular distributions are focused at too forward
an angle, as reported in Ref.~\cite{Cap12}.
This lack of Coulomb deflection of the eikonal approximation
can be solved using the empirical shift Eq.~(\ref{ebp}).
Using this shift the agreement with CDCC improves significantly.

In addition, E-CDCC turns out to have a convergence problem, which indicates
the limit of application of the eikonal approximation
using a discretized continuum
to reactions at such low energies involving a strong Coulomb interaction.
Fortunately, by including
a QM correction to E-CDCC for lower
projectile-target partial waves, this convergence problem is completely resolved.
Moreover, the result of this hybrid calculation perfectly agrees with the result
of the full (QM) CDCC.

The present study confirms the difficulty to properly describe
the Coulomb interaction within the eikonal approximation at low energy.
However, it is found that even at 20~MeV/nucleon,
the empirical shift Eq.~(\ref{ebp}) helps correctly reproducing the Coulomb
deflection that was shown to be missing in the DEA \cite{Cap12}.
Including QM corrections within the E-CDCC leads to a hybrid model
that exhibits the same accuracy as the full CDCC, with a minimal computational cost.
This hybrid calculation will be useful for describing nuclear and
Coulomb breakup processes in a wide range of incident energies.
It could be included in other CDCC programs to increase their computational
efficiency without reducing their accuracy.
This could be an asset to improve the description of projectiles while keeping
reasonable calculation times.

\section*{Acknowledgment}

The authors thank D.~Baye and M.~Yahiro for valuable comments on this study.
This research was supported in part by the Fonds de la Recherche Fondamentale
Collective (Grant No.~2.4604.07F),
Grant-in-Aid of the Japan Society for the Promotion of Science (JSPS),
and RCNP Young Foreign Scientist Promotion Program.
This paper presents research results of the Belgian Research Initiative on eXotic nuclei
(BRIX), Program No. P7/12, on interuniversity attraction poles of the
Belgian Federal Science Policy Office.

\end{document}